\documentclass[12pt,preprint,notoc,nohyper]{QCEMDA} % 10pt is ignored!

\usepackage{epsfig,multicol,bbm}

\newcommand\fverb{\setbox\pippobox=\hbox\bgroup\verb}
\newcommand\fverbdo{\egroup\medskip\noindent%
            \fbox{\unhbox\pippobox}\ }
\newcommand\fverbit{\egroup\item[\fbox{\unhbox\pippobox}]}
\newbox\pippobox
\title{Hawking temperature of rotating charged black strings from tunneling}

\author{Jamil Ahmed and K. Saifullah  \\

Department of Mathematics, Quaid-i-Azam University, Islamabad,
Pakistan \\

Electronic address: \email{saifullah@qau.edu.pk}}

\preprint{}  % OR: \preprint{Aaaa/Mm/Yy\\Aaa-aa/Nnnnnn}
\abstract{Thermal radiations from spherically symmetric black holes
have been studied from the point of view of quantum tunneling. In
this paper we extend this approach to study radiation of fermions
from charged and rotating black strings. Using WKB approximation and
Hamilton-Jacobi method we work out the tunneling probabilities of
incoming and outgoing fermions and find the correct Hawking
temperature for these objects. We show that in appropriate limits
the results reduce to those for the uncharged and non-rotating black
strings.}

\dedicated{Dedicated to Professor Asghar Qadir on his sixty fifth
birthday.}

\begin{document}

\section{Introduction}

%\cite{HT, AK, Ka, PL, DL} \cite{KM06, KM08a, KM08b, ZL, LRW, CJZ,Ji,
%ZY, DJ, YY, GS, RS, AS11a}

Cylindrically symmetric static solutions of the Einstein-Maxwell
equations with a negative cosmological constant represent charged
black strings. Soon after the pioneering work
\cite{lemos95a,lemos95b,S} on cylindrically symmetric black hole
solution its charged and rotating versions were found \cite{LZ, CZ}.
The rotating version is very similar to the Kerr-Newman black hole
in spherical symmetry. These black configurations have been studied
for different physical (\cite{HT}-\cite{DL}) and thermodynamical
\cite{De, FS} properties.

One of the important recent developments in theoretical physics is
studying Hawking radiation \cite{Hawk1} from the point of view of
quantum tunneling from black hole horizons \cite{PW, KW, Pa}. In the
so-called Hamilton-Jacobi method the tunneling rate of particles
coming out from the black hole horizon is calculated from the
imaginary part of the action \cite{SP, SPS}. This approach has been
successfully applied to various spherically and axially symmetric
black configurations (\cite{KM06}-\cite{AS11a}). In this paper we
extend the analysis to tunneling of fermions from anti-de Sitter
rotating charged black strings. We use WKB approximation to find the
tunneling probabilities of incoming and outgoing charged fermions,
and calculate the Hawing temperature. If the charge is taken to be
zero one can deduce the temperature for the uncharged rotating black
strings. Our results support the recent claim \cite{FS} that the
Hawking temperature for these strings as calculated earlier
\cite{lemos95b} is incorrect. Thus we have an instance where the
tunneling method confirms the correctness of the Hawking
temperature. This was indeed one of the aims and objectives of
developing this approach.

The paper is organized as follows. In Section 2 we describe the
spacetime for rotating charged black strings. Section 3 deals with
the solution of charged Dirac equation in the background of these
black strings and Hawking temperature is calculated. In Conclusion
we show how this temperature reduces to those in the simpler cases
of uncharged and non-rotating black strings.

\section{Rotating charged black string}

Analogous to the Kerr-Newman black hole solution in spherical
symmetry, a solution for rotating charged object with cylindrical
geometry in the presence of negative cosmological constant was found
\cite{LZ}. Both these solutions have event and Cauchy horizons,
closed timelike curves and timelike singularities. The solution for
the black string is given by the line element
\begin{eqnarray}
ds^{2} &=&-\left( \alpha ^{2}r^{2}-\frac{2G\left( M+\Omega \right)
}{\alpha r }+\frac{4GQ^{2}}{\alpha ^{2}r^{2}}\right)
dt^{2}-\frac{16GJ}{\alpha r}\left( 1-\frac{2Q^{2}}{\left( M+\Omega
\right) \alpha r}\right) dtd\theta  \nonumber
\\
&&+\left( r^{2}+\frac{4G\left( M-\Omega \right) }{\alpha
^{3}r}\left( 1- \frac{2Q^{2}}{\left( M-\Omega \right) \alpha
r}\right) \right) {\large
d\theta }^{2}+\alpha ^{2}r^{2}dz^{2}  \nonumber \\
&&+\frac{dr^{2}}{\alpha ^{2}r^{2}-2G\left( 3\Omega -M\right) /\alpha
r+4GQ^{2} \left( 3\Omega -M\right) /\left( M+\Omega \right) \alpha
^{2}r^{2}} .  \label{5.1}
\end{eqnarray}
Here
\begin{equation}
\Omega =\sqrt{M^{2}-\frac{8J^{2}\alpha ^{2}}{9}}.  \label{5.2}
\end{equation}
We define $a$ as
\begin{equation}
a^{2}\alpha ^{2}=1-\frac{\Omega }{M},  \label{5.3}
\end{equation}
such that
\begin{equation}
M+\Omega =2M\left(1-\frac{a^{2}\alpha^{2}}{2}\right) . \label{5.4}
\end{equation}
Thus metric (\ref{5.1}) can also be written as
\begin{eqnarray}
ds^{2} &=&-\left( \alpha ^{2}r^{2}-\frac{4M\left( 1-a^{2}\alpha
^{2}/2\right) }{\alpha r}+\frac{4Q^{2}}{\alpha ^{2}r^{2}}\right)
dt^{2}
\nonumber \\
&&-\frac{4aM\sqrt{1-a^{2}\alpha ^{2}/2}}{\alpha r}\left(
1-\frac{Q^{2}}{
M\left( 1-a^{2}\alpha ^{2}/2\right) \alpha r}\right) 2dtd\theta  \nonumber \\
&&+\left( \alpha ^{2}r^{2}-\frac{4M\left( 1-3a^{2}\alpha
^{2}/2\right) }{ \alpha r}+\frac{4Q^{2}}{\alpha ^{2}r^{2}}\left(
\frac{1-3a^{2}\alpha ^{2}/2}{
1-a^{2}\alpha ^{2}/2}\right) \right) ^{-1}dr^{2}  \nonumber \\
&&+\left( r^{2}+\frac{4Ma^{2}}{\alpha r}\left( 1-\frac{Q^{2}}{\left(
1-a^{2}\alpha ^{2}/2\right) M\alpha r}\right) \right) d\theta
^{2}+\alpha ^{2}r^{2}dz^{2}.  \label{5.5}
\end{eqnarray}
Here $-\infty <z<+\infty $, $0<\alpha z<2\pi $. Defining \cite{LZ}
\[
\Delta =\alpha ^{2}r^{2}-\frac{b}{\alpha r}+\frac{c^{2}}{\alpha
^{2}r^{2}} , \,\,\,\, b=4M\left( 1-\frac{3}{2}\alpha
^{2}a^{2}\right) ,
\]
\[
c^{2}=4Q^{2}\left( \frac{1-3\alpha ^{2}r^{2}/2}{1-\alpha
^{2}r^{2}/2}\right) , \,\,\, \gamma =\sqrt{\frac{1-a^{2}\alpha
^{2}/2}{1-3a^{2}\alpha ^{2}/2} } \,\,\, \textrm{and } \,\,\, \omega
=\frac{a\alpha ^{2}}{\sqrt{1-3a^{2}\alpha ^{2}/2}},
\]
the above metric takes the form
\begin{equation}
ds^{2}=-\Delta \left( \gamma dt-\frac{\omega }{\alpha ^{2}}d\theta
\right) ^{2}+r^{2}\left( \gamma d\theta -\omega dt\right)
^{2}+\frac{dr^{2}}{\Delta } +\alpha ^{2}r^{2}dz^{2}.  \label{5.6}
\end{equation}
We choose new angular coordinate given by $\overline{\theta }=\gamma
\theta -\omega t$ such that $d\theta =d\overline{\theta }/\gamma -
\omega dt/\gamma$. With this choice of the coordinate the above
metric becomes
\begin{eqnarray}
ds^{2}&=&-\left( \alpha ^{2}r^{2}-\frac{b}{\alpha
r}+\frac{c^{2}}{\alpha ^{2}r^{2}}\right) \left( \frac{dt}{\gamma
}-\frac{\omega d\overline{\theta } }{\alpha ^{2}\gamma }\right) ^{2}
\nonumber \\
&+& r^{2}d\overline{\theta }+\left( \alpha ^{2}r^{2}-\frac{b}{\alpha
r}+\frac{c^{2}}{\alpha ^{2}r^{2}}\right) ^{-1}dr^{2}+\alpha
^{2}r^{2}dz^{2}.  \label{5.7}
\end{eqnarray}
The electromagnetic potential for the black string is given by
\cite{LZ}
\[
A_{0}=-\gamma h\left( r\right) , \,\,\, A_{2}=\frac{\omega }{\alpha
^{2}} h\left( r\right) , \,\,\, A_{1}=0=A_{3}.
\]
Here $h\left( r\right) =2\lambda /\alpha r$, where $\omega$ and
$\gamma $ are constants. In order to obtain the horizon of the black
hole we put $g^{11}=0$, so that
\[
\alpha ^{2}r^{2}-\frac{b}{\alpha r}+\frac{c^{2}}{\alpha
^{2}r^{2}}=0,
\]
which on solving yields
\begin{equation}
r_{\pm }=\frac{b^{1/3}\sqrt{s}\pm
\sqrt{2\sqrt{s^{2}-4q^{2}}-s}}{2\alpha }. \label{5.8}
\end{equation}
Here
\begin{equation}
s=\left( \frac{1}{2}+\frac{1}{2}\sqrt{1-4\left(
\frac{4q^{2}}{3}\right) ^{3}} \right) ^{1/3}+\left(
\frac{1}{2}-\frac{1}{2}\sqrt{1-4\left( \frac{4q^{2}}{3} \right)
^{3}}\right) ^{1/3},  \label{5.9}
\end{equation}
where $q^{2}=c^{2}/b^{4/3}$. We put $F(r)=-(g^{tt})^{-1}$, so that
\begin{equation}
F(r)=\frac{-\Delta \gamma^{2}
\alpha^4}{\gamma^{4}\alpha^4+\omega^{2}\Delta}  ,
\end{equation}
and
\begin{equation}
F_{r}(r_+)=\frac{\Delta^{\prime}}{\gamma ^{2}} .   \label{5.10}
\end{equation}
We also let
\begin{equation}
g(r)=\Delta ,
\end{equation}
so that
\begin{equation}
g_{r}(r_+)=\Delta^{\prime} .  \label{5.11}
\end{equation}
This notation will be used later, and prime here denotes a
derivative with respect to $r$.

\section{Tunneling probability and Hawking temperature}

In this section we work out the probability of fermions to tunnel
across the event horizon of rotating charged black string. In order
to do this we solve the Dirac equation using WKB approximation. An
important outcome of this procedure is the computation of correct
Hawking temperature for these black objects. The charged Dirac
equation for the field $\Psi$, and mass, $m$, and charge, $q$, of
the particle is
\begin{equation}
\iota \gamma ^{\mu }\left( D_{\mu }-\frac{\iota q}{\hbar }A_{_{\mu
}}\right) \Psi +\frac{m}{\hbar }\Psi =0.  \label{5.12}
\end{equation}%
Here%
\[
D_{\mu }=\partial _{\mu }+\Omega _{\mu }.
\]%
Also%
\[
\Omega _{\mu }=\frac{1}{2}\iota \Gamma _{\mu }^{\alpha \beta }\Sigma
_{\alpha \beta },\Sigma _{\alpha \beta }=\frac{1}{4}\iota \left[
\gamma ^{\alpha },\gamma ^{\beta }\right] .
\]%
Using properties of the commutative brackets we note that $\Omega
_{\mu }=0$, so that
\[
D_{\mu }=\partial _{\mu },
\]%
and Eq. (\ref{5.12}) becomes
\begin{equation}
\iota \gamma ^{\mu }\left( \partial _{\mu }-\frac{\iota q}{\hbar
}A_{_{\mu }}\right) \Psi +\frac{m}{\hbar }\Psi =0.  \label{5.13}
\end{equation}
Applying summation this takes the form
\begin{equation}
\iota \gamma ^{t}\partial _{t}\Psi +\iota \gamma ^{r}\partial
_{r}\Psi +\iota \gamma ^{\theta }\partial _{\theta }\Psi +\iota
\gamma ^{z}\partial _{z}\Psi +\frac{q}{\hbar }A_{t}\gamma ^{t}\Psi
+\frac{q}{\hbar }A_{\theta }\gamma ^{\theta }\Psi +\frac{m}{\hbar
}\Psi =0. \label{5.14}
\end{equation}
We choose the following $\gamma$-matrices
\[
\gamma ^{t}=\frac{1}{\sqrt{F\left( r\right) }}\left(
\begin{array}{cc}
0 & I \\
-I & 0
\end{array}
\right) , \gamma ^{r}=\sqrt{g\left( r\right) }\left(
\begin{array}{cc}
0 & \sigma ^{3} \\
\sigma ^{3} & 0
\end{array}
\right),
\]
\[
\gamma ^{\theta }=M\left( r\right) \left(
\begin{array}{cc}
0 & \sigma ^{2} \\
\sigma ^{2} & 0
\end{array}
\right) +M\left( r\right) N\left( r\right) \left(
\begin{array}{cc}
0 & I \\
-I & 0
\end{array}
\right) ,
\]
\[
\gamma ^{z}=\frac{1}{\alpha r}\left(
\begin{array}{cc}
0 & \sigma ^{1} \\
\sigma ^{1} & 0
\end{array}
\right) ,
\]
where $\sigma ^{i}$ are Pauli matrices, and $M\left( r\right) $ and
$N\left( r\right) $ are defined as
\[
M\left( r\right) =\frac{\alpha ^{2}\gamma }{\omega} \left(
\frac{\omega ^{2}-\Delta}{r^{2} \alpha^{4}\gamma^{2}}\right)^{1/2},
\]
\[
N\left( r\right) =\frac{\gamma }{\Delta^{1/2}}.
\]

We assume the following form of Dirac's field for Eq. (\ref{5.14})
\begin{equation}
\Psi\left( t,r,\theta ,z\right) =\left(
\begin{array}{c}
A\left( t,r,\theta ,z\right) \xi \\
B\left( t,r,\theta ,z\right) \xi
\end{array}
\right) \exp \left( \frac{\iota }{\hbar }I\right) ,  \label{3.17}
\end{equation}
where $I$ is the classical action, and $A$ and $B$ are arbitrary
functions of the coordinates. Substituting this in Eq. (\ref{5.14})
we apply WKB approximation, divide by the exponential term and
multiply by $\hbar$. Thus the resulting equations to leading order
in $\hbar $ take the form
\begin{equation}
-B\left( \frac{I_{t}}{\sqrt{F\left( r\right) }}+\sqrt{g\left(
r\right) } I_{r}-M\left( r\right) N\left( r\right) I_{\theta
}-\frac{qA_{t}}{\sqrt{ F\left( r\right) }}-qA_{\theta }M\left(
r\right) N\left( r\right) \right) +mA=0,  \label{5.23}
\end{equation}

\begin{equation}
A\left( \frac{I_{t}}{\sqrt{F\left( r\right) }}-\sqrt{g\left(
r\right) } I_{r}-M\left( r\right) N\left( r\right) I_{\theta
}-\frac{qA_{t}}{\sqrt{ F\left( r\right) }}-qA_{\theta }M\left(
r\right) N\left( r\right) \right)+mB=0.  \label{5.24}
\end{equation}
Considering the Killing vectors of the background spacetime we
employ the following ansatz
\begin{equation}
I\left( t,r,\theta ,z\right) =-Et+l\theta +Jz+W\left( r\right) ,
\label{5.25}
\end{equation}
where $E$ is the energy of the emitted particles and $W$ is the part
of the action $I$ that contributes to the tunneling probability.
Using Eq. (\ref{5.25}) we find that Eqs. (\ref{5.23}) and
(\ref{5.24}) become
\begin{equation}
B\left( \frac{E}{\sqrt{F\left( r\right) }}-\sqrt{g\left( r\right)
}\frac{dW }{dr}+M\left( r\right) N\left( r\right)
l+\frac{qA_{t}}{\sqrt{F\left( r\right) }}+qA_{\theta }M\left(
r\right) N\left( r\right) \right) \nonumber \\ +mA=0,  \label{5.26}
\end{equation}

\begin{equation}
A\left( \frac{-E}{\sqrt{F\left( r\right) }}-\sqrt{g\left( r\right)
}\frac{dW }{dr}-M\left( r\right) N\left( r\right)
l-\frac{qA_{t}}{\sqrt{F\left( r\right) }}-qA_{\theta }M\left(
r\right) N\left( r\right) \right) \nonumber \\ +mB=0.  \label{5.27}
\end{equation}
We first calculate the function $W(r)$ for the massless case, i.e.,
for $m=0$. From the above equations we can write
\[
\frac{dW}{dr}=\frac{E+M\left( r\right) N\left( r\right)
\sqrt{F\left( r\right) }l-qA_{t}-qA_{\theta }M\left( r\right)
N\left( r\right) \sqrt{ F\left( r\right) }}{\sqrt{F\left( r\right)
g\left( r\right) }},
\]
or
\begin{equation}
W_{+}\left( r\right) =\int \frac{E+M\left( r\right) N\left( r\right)
\sqrt{ F\left( r\right) }l-qA_{t}-qA_{\theta }M\left( r\right)
N\left( r\right) \sqrt{F\left( r\right) }}{\sqrt{F\left( r\right)
g\left( r\right) }}dr. \label{5.28}
\end{equation}
Now expanding $F\left( r\right) $ and $g\left( r\right) $ in
Taylor's series near the horizon and neglecting squares and higher
powers of $\left( r-r_{+}\right) $ we get
\begin{equation}
g\left( r\right) =g\left( r_{+}\right) +\left( r-r_{+}\right)
\partial _{r}g\left( r_{+}\right) ,  \label{5.29}
\end{equation}
which can be written as
\[
g(r) =( r-r_{+}) \Delta^{\prime}  .
\]
Also
\[
F(r) =\frac{(r-r_{+})}{\gamma^2} \Delta^{\prime}  .
\]
Using these values in Eq. (\ref{5.28}) we obtain%
\[
W\left( r_{+}\right) =\int \frac{E+M\left( r\right) N\left( r\right)
\sqrt{ F\left( r\right) }l-qA_{t}-qA_{\theta }M\left( r\right)
N\left( r\right) \sqrt{F\left( r\right) }}{\left( r-r_{+}\right)
\frac{1}{\gamma }\left( 2\alpha ^{2}r_{+}+\frac{b}{\alpha
r_{+}^{2}}-\frac{2c^{2}}{\alpha ^{2}r_{+}^{3}}\right) }dr.
\]%
Integrating around the simple pole we get
\begin{equation}
W_{+}=\pi \iota \gamma \left( \frac{E+M\left( r\right) N\left(
r\right) \sqrt{F\left( r\right) }l-qA_{t}-qA_{\theta }M\left(
r\right) N\left( r\right) \sqrt{F\left( r\right) }}{\left( 2\alpha
^{2}r_{+}+\frac{b}{\alpha r_{+}^{2}}-\frac{2c^{2}}{\alpha
^{2}r_{+}^{3}}\right) }\right) . \label{5.29a}
\end{equation}
Similarly
\begin{equation}
W_{-}=\pi \iota \gamma \left( \frac{-E+M\left( r\right) N\left(
r\right) \sqrt{F\left( r\right) }l-qA_{t}-qA_{\theta }M\left(
r\right) N\left( r\right) \sqrt{F\left( r\right) }}{\left( 2\alpha
^{2}r_{+}+\frac{b}{\alpha r_{+}^{2}}-\frac{2c^{2}}{\alpha
^{2}r_{+}^{3}}\right) }\right) . \label{5.29b}
\end{equation}
The probabilities of fermions crossing the horizon in each direction
are
\[
P_{emission} \varpropto \exp \left( -2 Im I\right) =\exp \left( -2
Im W_{+}\right)
\]%

\[
P_{absorption} \varpropto \exp \left( -2 Im I\right) =\exp \left( -2
Im W_{-} \right) .
\]%
While computing the imaginary part of the action, we note that it is
same both for the incoming and outgoing solutions, and so will
cancel out. Now the probability of particles tunneling from inside
to outside the horizon is given by \cite{SP,SPS}
\[
\Gamma \varpropto \frac{P_{emission}}{P_{absorption}}=\frac{ \exp
\left( -2Im W_{+}\right) }{\exp \left( -2 Im W_{-}\right)},
\]
or

\begin{equation}
\Gamma =\exp \left( -4Im W_{+}\right) . \label{5.30}
\end{equation}
Using the value of $W_{+}$ in this we obtain
\begin{equation}
\Gamma =\exp \left[ -4\pi \gamma \left( \frac{E+M\left( r\right)
N\left( r\right) \sqrt{F\left( r\right) }l-qA_{t}-qA_{\theta
}M\left( r\right)
N\left( r\right) \sqrt{F\left( r\right) }}{\left( 2\alpha ^{2}r_{+}+\frac{b}{%
\alpha r_{+}^{2}}-\frac{2c^{2}}{\alpha ^{2}r_{+}^{3}}\right)
}\right) \right] . \label{5.31}
\end{equation}
Here we note that the probabilities do not violate unitarity by
exceeding the value of 1. This is because besides the spatial
contribution there is a temporal part also which contributes both to
emission and absorption prbabilities \cite{APGS, APS, AAS} and we
obtain a correct value of $\Gamma$. When we compare this with
$\Gamma = \exp \left( -\beta E\right)$, where $\beta =1/T_{H}$, we
note that

\begin{equation}
\beta =\frac{-4\pi \gamma }{2\alpha ^{2}r_{+}+\frac{b}{\alpha r_{+}^{2}}-%
\frac{2c^{2}}{\alpha ^{2}r_{+}^{3}}},   \label{5.33}
\end{equation}
so that the Hawking temperature for rotating charged black strings
comes out to be
\begin{equation}
T_{H}=\frac{1}{4\pi \gamma }\left( 2\alpha ^{2}r_{+}+\frac{b}{\alpha
r_{+}^{2}}-\frac{2c^{2}}{\alpha ^{2}r_{+}^{3}}\right) , \label{5.34}
\end{equation}
Using the values of $\gamma ,$ $b$ and $c^{2}$ this takes the form
\begin{equation}
T=\frac{1}{2\pi }\sqrt{\frac{1-\frac{1}{2}\alpha
^{2}a^{2}}{1-\frac{3}{2} \alpha ^{2}a^{2}}}\left[ \alpha
^{2}r_{+}+\frac{2M\left( 1-\frac{3}{2}\alpha ^{2}a^{2}\right)
}{\alpha ^{2}r_{+}^{2}}-\frac{4Q^{2}}{\alpha ^{2}r_{+}^{3}} \left(
\frac{1-\frac{3}{2}\alpha ^{2}a^{2}}{1-\frac{1}{2}\alpha ^{2}a^{2}}
\right) \right] .  \label{ht}
\end{equation}
To recover Hawking temperature for the non-rotating case, we put
$a=0$ to get

\begin{equation}
T_{H}=\frac{1}{2\pi }\left( \alpha ^{2}r_{+}+\frac{2M}{\alpha
r_{+}^{2}}- \frac{4Q^{2}}{\alpha ^{2}r_{+}^{3}}\right) ,
\end{equation}
which is in agreement with the result derived in Refs.
\cite{CZ,FS,AS11a}. For the massive case we multiply Eq.
(\ref{5.26}) by $A$ and Eq. (\ref{5.27}) by $B$ to obtain

\begin{eqnarray}
AB\left( E-\sqrt{g\left( r\right) F(r)}\frac{dW}{dr}+M\left(
r\right) N\left( r\right) l\sqrt{F\left( r\right)
}+qA_{t}+qA_{\theta }M\left( r\right) N\left( r\right) \sqrt{F\left(
r\right) }\right) \nonumber \\ +  \sqrt{F\left( r\right) }mA^{2}=0 ,  \nonumber \\
\label{5.35} \\
AB\left( -E-\sqrt{g\left( r\right) F(r)}\frac{dW}{dr}-M\left(
r\right) N\left( r\right) l\sqrt{F\left( r\right)
}-qA_{t}-qA_{\theta }M\left( r\right) N\left( r\right) \sqrt{F\left(
r\right) }\right) \nonumber  \\ + \sqrt{F\left( r\right) }mB^{2}=0 .
\nonumber \\  \label{5.36}
\end{eqnarray}
Subtracting Eq. (\ref{5.36}) from Eq. (\ref{5.35}) gives

\begin{eqnarray}
\frac{A}{B} &=&\frac{1}{m\sqrt{F\left( r\right) }}\left( \left(
E+M\left( r\right) N\left( r\right) l\sqrt{F\left( r\right)
}+qA_{t}+qA_{\theta }M\left( r\right) N\left( r\right) \sqrt{F\left(
r\right) }\right) \pm
\right.  \nonumber \\
&&\left. \sqrt{\left( E+M\left( r\right) N\left( r\right)
l\sqrt{F\left( r\right) }+qA_{t}+qA_{\theta }M\left( r\right)
N\left( r\right) \sqrt{ F\left( r\right) }\right) ^{2}+m^{2}F\left(
r\right) }\right)  .\nonumber \\ \label{5.37}
\end{eqnarray}
We see that
\[
\lim_{r\rightarrow r_{+}}\left( \frac{A}{B}\right) =\left\{
\begin{array}{c}
0 \,\,\ \textrm{for upper sign}\left( +\right) \\
-\infty \,\,\  \textrm{for lower sign}\left( -\right)
\end{array}
\right. .
\]
At the horizon either $A=0$ or $B=0.$ For $A\rightarrow 0,$ we solve
Eq. (\ref{5.27}) for $m$ and insert in Eq. (\ref{5.26}) to get after
simplification

\begin{equation}
\left( \frac{dW_{+}}{dr}\right) =\frac{E+M\left( r\right) N\left(
r\right) +lqA_{t}+qA_{\theta }M\left( r\right) N\left( r\right)
}{\sqrt{F_{r}\left( r\right) g_{r}\left( r\right) }\left(
r-r_{+}\right) }\left( \frac{ 1+A^{2}/B^{2}}{1-A^{2}/B^{2}}\right) .
\label{5.38}
\end{equation}
Similarly, for $B\rightarrow 0$ we have
\begin{equation}
\left( \frac{dW_{-}}{dr}\right) =\frac{- E+M\left( r\right) N\left(
r\right) +lqA_{t}+qA_{\theta }M\left( r\right) N\left( r\right)
 }{ \sqrt{F_{r}\left( r\right) g_{r}\left( r\right) }\left(
r-r_{+}\right) } \left( \frac{1+A^{2}/B^{2}}{1-A^{2}/B^{2}}\right) .
\label{5.39}
\end{equation}
Integrating Eq. (\ref{5.38}) around the contour and using the values
of $F_{r}\left( r\right) $ and $g_{r}\left( r\right) $ from Eqs.
(\ref{5.10}) and (\ref{5.11}) yields
\begin{equation}
W_{+}=\frac{\pi \iota \gamma \left( E+M\left( r\right) N\left(
r\right) +lqA_{t}+qA_{\theta }M\left( r\right) N\left( r\right)
\right) }{\left( 2\alpha ^{2}r_{+}+\frac{b}{\alpha
r_{+}^{2}}-\frac{2c^{2}}{\alpha ^{2}r_{+}^{3}}\right) }.
\label{5.40}
\end{equation}
Similarly, from Eq. (\ref{5.39}) we get
\begin{equation}
W_{-}=\frac{\pi \iota \gamma \left( -E+M\left( r\right) N\left(
r\right) +lqA_{t}+qA_{\theta }M\left( r\right) N\left( r\right)
\right) }{\left( 2\alpha ^{2}r_{+}+\frac{b}{\alpha
r_{+}^{2}}-\frac{2c^{2}}{\alpha ^{2}r_{+}^{3}}\right) }.
\label{5.41}
\end{equation}
Thus in this case we obtain
\[
\Gamma =-4\pi \left( \frac{\gamma \left( E+M\left( r\right) N\left(
r\right) +lqA_{t}+qA_{\theta }M\left( r\right) N\left( r\right)
\right) }{\left( 2\alpha ^{2}r_{+}+\frac{b}{\alpha
r_{+}^{2}}-\frac{2c^{2}}{\alpha ^{2}r_{+}^{3}}\right) }\right) .
\]
This gives the Hawking temperature as
\[
T_{H}=\frac{1}{4\pi \gamma }\left( 2\alpha ^{2}r_{+}+\frac{b}{\alpha
r_{+}^{2}}-\frac{2c^{2}}{\alpha ^{2}r_{+}^{3}}\right) .
\]%
Using the values of $\gamma $, $b$ and $c^{2}$ we get the same
result as in Eq. (\ref{ht}).

\section{Discussion}

Hawking radiation can be viewed as a process of quantum tunneling
from black hole horizons by using the techniques of quantum field
theory on curved spacetimes. Mathematically this is achieved by the
Hamilton-Jacobi method by using WKB approximation and complex path
integration. The particles traverse trajectories which are forbidden
classically. These paths are given in terms of the imaginary part of
their classical action.

We have considered cylindrically symmetric solutions of the
Einstein-Maxwell equations which are anti-de Sitter type. We have
studied radiation of charged fermions from these black strings. The
method yields the tunneling probability of the particles crossing
the horizon, and the Hawking temperature for rotating charged black
string which is given by

\begin{equation}
T=\frac{1}{2\pi }\sqrt{\frac{1-\frac{1}{2}\alpha ^{2}a^{2}}{1-\frac{3}{2}%
\alpha ^{2}a^{2}}}\left[ \alpha ^{2}r_{+}+\frac{2M\left(
1-\frac{3}{2}\alpha
^{2}a^{2}\right) }{\alpha ^{2}r_{+}^{2}}-\frac{4Q^{2}}{\alpha ^{2}r_{+}^{3}}%
\left( \frac{1-\frac{3}{2}\alpha ^{2}a^{2}}{1-\frac{1}{2}\alpha ^{2}a^{2}}%
\right) \right] .  \label{5.3}
\end{equation}
We see that from this formula we can recover the temperature for
\emph{simpler} cases. For example, if we put charge, $Q=0$, we
obtain the result for the uncharged rotating black string as
\begin{equation}
T=\frac{1}{2\pi }\sqrt{\frac{1-\frac{1}{2}\alpha ^{2}a^{2}}{1-\frac{3}{2}%
\alpha ^{2}a^{2}}}\left[ \alpha ^{2}r_{+}+\frac{2M\left(
1-\frac{3}{2}\alpha ^{2}a^{2}\right) }{\alpha ^{2}r_{+}^{2}}\right]
.  \label{5.4}
\end{equation}
It has been noted recently \cite{FS} that the temperature for this
case as given earlier \cite{lemos95b} is incorrect. Here it is
confirmed by the tunneling procedure, and this is one of the
successes of this method that it provides the correct Hawking
temperature. This was also an important motivation behind developing
this approach.

Further, putting $a=0$ in Eq. (\ref{5.4}) yields temperature for the
non-rotating case as
\begin{equation}
T=\frac{1}{2\pi }\left( \alpha ^{2}r_{+}+\frac{2M}{\alpha r_{+}^{2}}-\frac{%
4Q^{2}}{\alpha ^{2}r_{+}^{3}}\right) ,  \label{5.5}
\end{equation}
which is consistent with the literature \cite{CZ,FS,AS11a}. Taking
$Q=0=a$ recovers the formula for non-rotating uncharged black string
\begin{equation}
T=\frac{1}{2\pi }\left( \alpha ^{2}r_{+}+\frac{2M}{\alpha
r_{+}^{2}}\right) . \label{5.6}
\end{equation}

\acknowledgments

We are thankful to Douglas Singleton for his comments.

\end{document}